\documentclass{epl}

\begin{document}

\title{From $ {\rm \bf VO_2} $ to $ {\rm \bf V_2O_3} $: The Metal-Insulator 
       Transition of the Magn\'{e}li Phase $ {\rm \bf V_6O_{11}} $}
\shorttitle{The Metal-Insulator Transition of $ {\rm V_6O_{11}} $}

\author{U.\ Schwingenschl\"ogl
        \thanks{E-mail: \email{Udo.Schwingenschloegl@physik.uni-augsburg.de}}
        \and V.\ Eyert \and U.\ Eckern}
\shortauthor{U.\ Schwingenschl\"ogl \etal}

\institute{Institut f\"ur Physik, Universit\"at Augsburg, 
           86135 Augsburg, Germany}


\maketitle

\begin{abstract}
The metal-insulator transition (MIT) of $ {\rm V_6O_{11}} $ is studied by 
means of electronic structure calculations using the augmented spherical 
wave method. The calculations are based on density functional theory 
and the local density approximation. Changes of the electronic 
structure at the MIT are discussed in relation to the structural 
transformations occuring simultaneously. The analysis is based on a 
unified point of view of the crystal structures of $ {\rm V_6O_{11}} $, 
$ {\rm VO_2} $, and $ {\rm V_2O_3} $. This allows to group the 
electronic bands into states behaving similar to the dioxide or the 
sesquioxide. While the sesquioxide-like V $ 3d_{yz} $ states show rather 
weak changes on entering the low-temperature structure, some of the 
dioxide-like V $ 3d_{x^2-y^2} $ states display splittings and shifts 
similar to those known from $ {\rm VO_2} $. The MIT of $ {\rm V_6O_{11}} $ 
arises as a combination of changes appearing in both of these compounds. 
Our results shed new light onto the role of particular electronic states 
for the MIT of $ {\rm V_2O_3} $. 
\end{abstract}

The binary oxides of vanadium are attracting considerable interest for many
years. This is due to metal-insulator transitions (MIT) displayed by the 
majority of these compounds, in particular their sensitivity to both 
electronic correlations and electron-phonon coupling. The delicate interplay 
of both mechanisms prevents a straightforward understanding of the 
transitions. As a consequence, the origins of the transitions are still a 
matter of dispute. This holds especially for the prototypical compounds 
$ {\rm VO_2} $ and $ {\rm V_2O_3} $\cite{goodenough71,brueckner83,imada98}. 

The MIT of $ {\rm VO_2} $ at 340\,K is accompanied by a structural 
transformation from the rutile structure to a monoclinic structure. 
While both electron-lattice interaction and electronic correlations 
have been proposed as the origin of the transition 
\cite{goodenough71,zylbersztejn75,wentzcovitch94}, recent electronic 
structure calculations for $ {\rm VO_2} $ gave strong hints at a Peierls 
instability of the one-dimensional $ d_{\parallel} $ ($ d_{x^2-y^2} $) 
band in an embedding background of the remaining V $ 3d $ $ t_{2g} $ states 
\cite{wentzcovitch94,habil}. This scenario was supported by studies of 
the neighbouring compounds $ {\rm MoO_2} $ and $ {\rm NbO_2} $ 
\cite{moo2pap,nbo2pap}. Yet, since calculations based on the local density
approximation (LDA) just miss the insulating gap of low-temperature 
$ {\rm VO_2} $, electronic correlations are also expected to be of some 
relevance. 

At a temperature of 168\,K and ambient pressure, stoichiometric 
$ {\rm V_2O_3} $ experiences a transition from a paramagnetic metallic 
(PM) phase to an antiferromagnetic insulating (AFI) phase. At the same 
time, the crystal structure transforms from the corundum to a monoclinic 
structure. In contrast, on doping with only small amounts of Al or Cr, 
$ {\rm V_2O_3} $ enters a paramagnetic insulating (PI) phase without 
change of the crystal symmetry. Yet, recent EXAFS experiments gave 
evidence of local crystal structure distortions in the PI phase, which 
only in the monoclinic AFI phase display long-range order 
\cite{pfalzer02}. The metal-insulator transitions of $ {\rm V_2O_3} $ are 
generally regarded as classical examples of a Mott-Hubbard transition 
\cite{castellani78}. LDA calculations performed for all three phases 
showed only a minor response of the electronic structures to the crystal 
parameter changes. The slight narrowing of the characteristic $ a_{1g} $ bands 
in the insulating phases formed the basis for a successful description of the 
PM-PI transition in a recent study, where LDA calculations were combined 
with dynamical mean-field theory (DMFT) \cite{held01}. This approach, called 
LDA+DMFT, is especially suited for a realistic modelling of strongly 
interacting electron systems. The strong influence of electronic correlations 
has also been demonstrated by LDA+U calculations for the AFI phase 
\cite{ezhov99}. 

A more comprehensive understanding of both compounds calls for deeper 
insight into the role of electronic correlations and electron-lattice 
interaction on going from $ {\rm VO_2} $ to $ {\rm V_2O_3} $. In 
particular, we are interested in the electronic states which are 
involved in the MIT. In the present paper we address these issues by 
discussing the results of electronic structure calculations for the 
Magn\'{e}li phase $ {\rm V_6O_{11}} $. The Magn\'{e}li phases of 
vanadium form a homologous series 
$ {\rm V}_n{\rm O}_{2n-1} $ ($ 3 \leq n \leq 9 $) and are particularly 
suited for studying the differences in crystal structures and electronic 
properties between the end members $ {\rm VO_2} $ ($ n \to \infty $) 
and $ {\rm V_2O_3} $ ($ n = 2 $). Being part of a broad investigation of the 
Magn\'{e}li phases, our work is the first {\em ab initio} study of these 
compounds at all. Relating the different local environments of the vanadium 
atoms in $ {\rm V_6O_{11}} $ to the electronic properties, we are able to 
obtain deeper insight into the crossover from the dioxide to the sesquioxide.
In particular, we find that the changes of the $ d_{x^2-y^2} $ states along 
the series account for most of the differences. 

As the general formula 
$ {\rm V}_n{\rm O}_{2n-1} = {\rm V_2O_3} + (n-2) {\rm VO_2} $ suggests,  
the crystal structures of the Magn\'{e}li phases are usually viewed as 
rutile-type slabs of infinite extension, which are separated by shear 
planes with a corundum-like atomic arrangement 
\cite{andersson63,brueckner83,horiuchi76}. While in the rutile-type 
regions the characteristic $ {\rm VO_6} $ octahedra are coupled via 
edges, the shear planes have face-sharing octahedra as in the corundum 
structure. 

\begin{figure}
\twoimages[width=65mm]{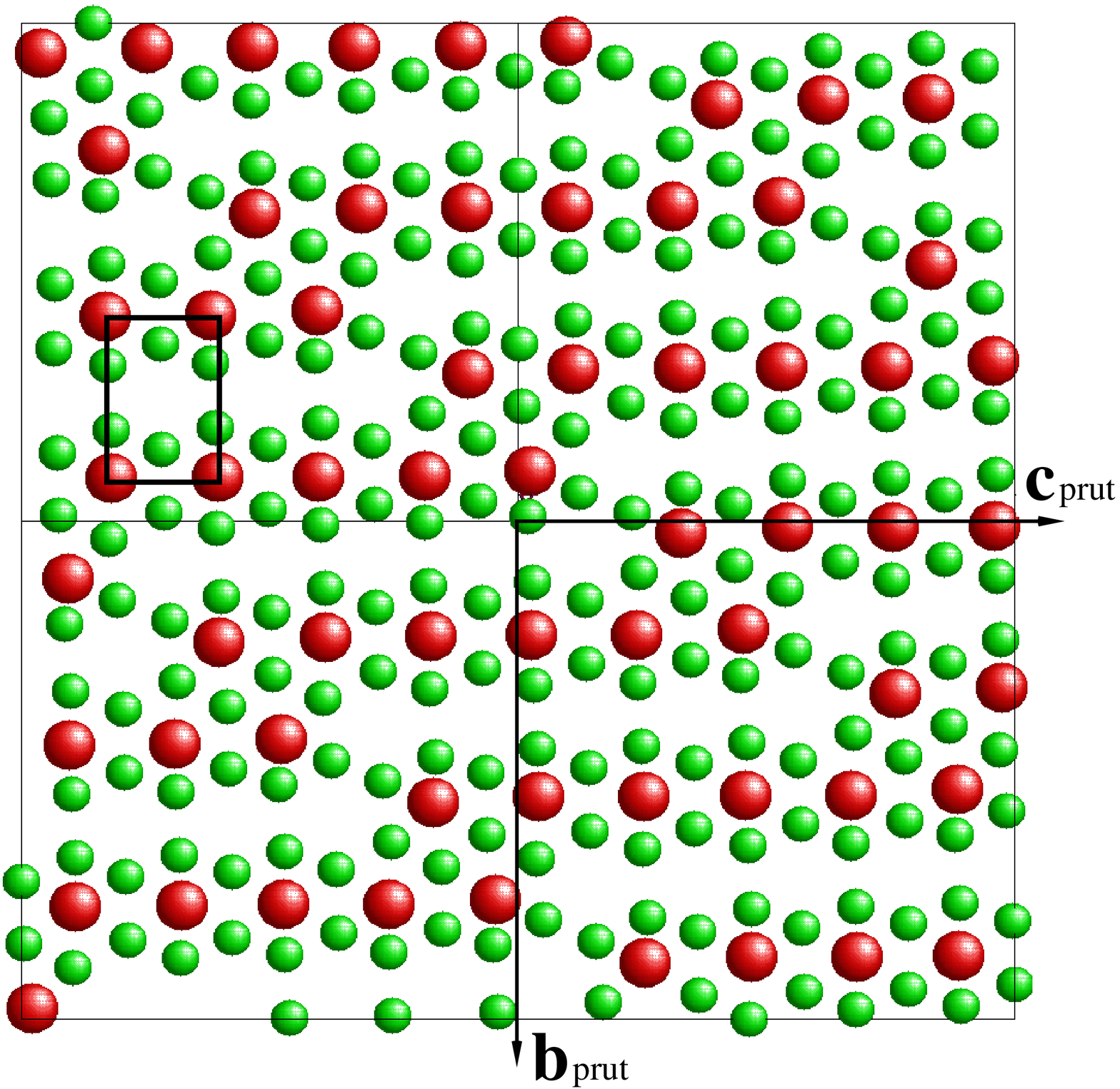}{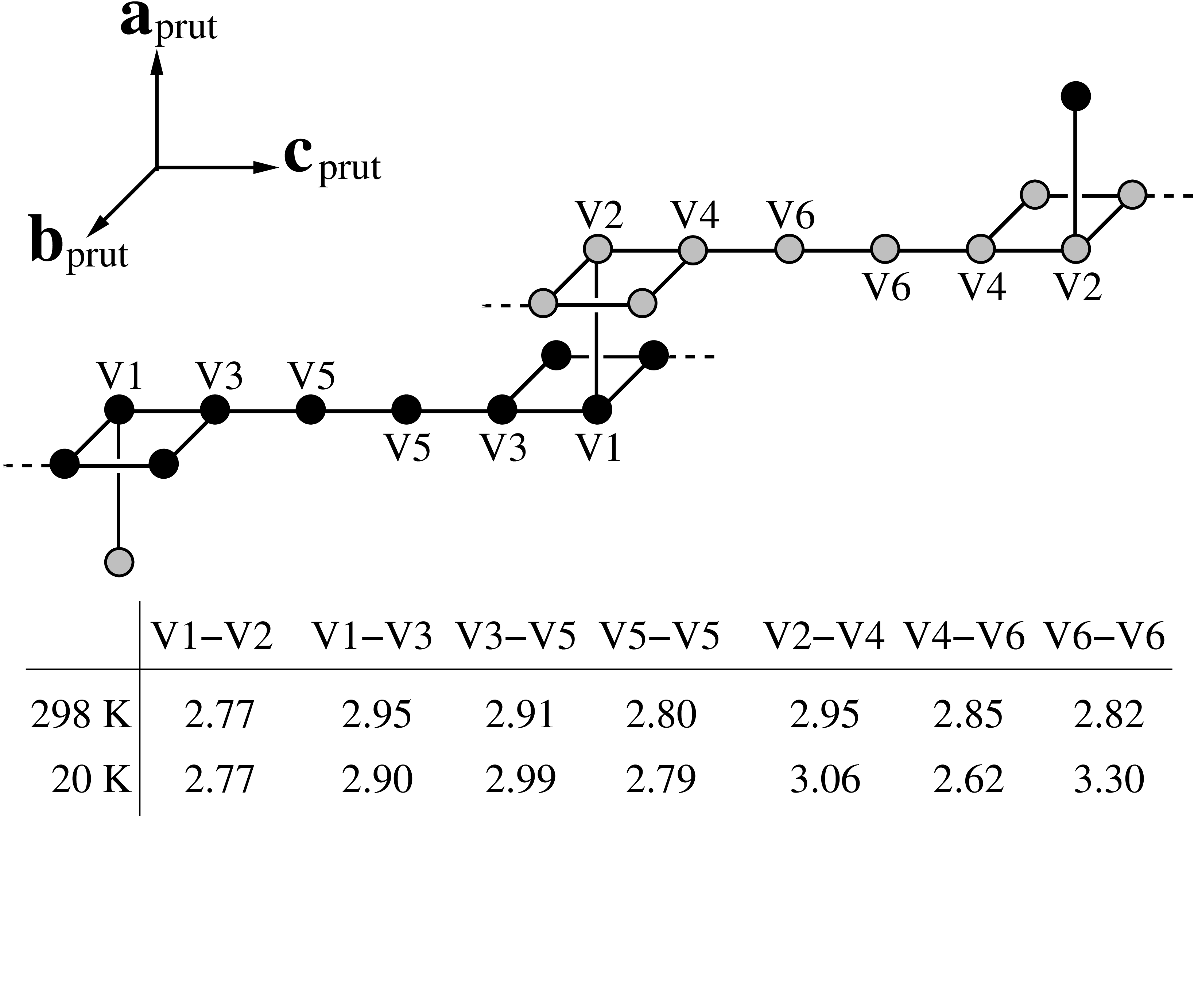}
\caption{Crystal structure of $ {\rm V_6O_{11}} $.
         Left: View along $ a_{\rm prut} $. Large and small circles mark 
         vanadium and oxygen atoms, respectively.
         Right: V-chains V1-V3-V5-V5-V3-V1 (black circles) and 
	 V2-V4-V6-V6-V4-V2 (gray circles) belonging to adjacent vanadium 
	 layers. The table lists measured V-V-distances (\AA) \protect 
         \cite{canfield90}.}
\label{fig1}
\end{figure}

For a deeper understanding of the electronic properties of all 
Magn\'{e}li phases we have developed a different representation 
of the crystal structures. It starts out from the regular 3D network of 
oxygen octahedra, which, apart from a slightly different buckling, is 
similar for all members including $ {\rm VO_2} $ and $ {\rm V_2O_3} $. 
Differences between the compounds $ {\rm V}_n{\rm O}_{2n-1} $ arise from 
the filling of the octahedra with vanadium atoms. Filled octahedra form 
chains of length $ n $ parallel to the pseudo-rutile axis 
$ c_{\rm prut} $, followed by $ n-1 $ empty sites. In particular, 
in $ {\rm VO_2} $ these chains have infinite length. For 
$ {\rm V_6O_{11}} $ the situation is sketched in Fig.~\ref{fig1}, 
which shows a projection of an O--V--O sandwich-like slab cut out of the 
crystal structure. In this projection, oxygen octahedra appear as 
a regular hexagonal network. Both this network and the vanadium chains 
of length 6 can be identified in the left panel of Fig.\ \ref{fig1}. 
Perpendicular to the projection, along $ a_{\rm prut} $, vanadium and 
oxygen layers alternate. Two different types of vanadium layers are 
distinguished, which comprise vanadium atoms V1, V3, V5, and V2, V4, V6, 
respectively. Within the vanadium sublattice these layers alternate along 
$ a_{\rm prut} $. However, vanadium chains in neighbouring vanadium layers 
are shifted parallel to $ c_{\rm prut} $, such that atoms V1 and V2 
are found on top of each other. This situation is sketched on the right 
panel of Fig.\ \ref{fig1}. The relative shifts of the vanadium chains 
give rise to different arrangements of the $ {\rm VO_6} $ octahedra. 
While octahedra neighbouring along $ a_{\rm prut} $ or $ b_{\rm prut} $ 
share faces, coupling between metal atoms along all other directions 
within the layers is via octahdral edges. Whereas the atomic arrangements 
near the ends of the chains are similar to those in $ {\rm V_2O_3} $, 
coordination near the chain centers is the same as in the rutile 
structure. The projection of the latter is indicated by a rectangle. 
Our representation offers the great advantage of allowing reference of the 
single symmetry components of the V $ 3d $ orbitals to coordinate systems, 
which are common to the whole class of compounds. Inspired by previous work 
on the dioxides we will use local coordinate systems with the $ z $ and 
$ x $ axis parallel to the apical axis of the local octahedron and the 
pseudo-rutile $ c_{\rm prut} $ axis, respectively \cite{habil,moo2pap,nbo2pap}. 

The MIT of $ {\rm V_6O_{11}} $ at 170\,K is accompanied by a structural 
transformation, which preserves the triclinic space group $ P\bar{1} $. 
Major changes concern the strong V4--V6 dimerization as well as 
ferroelectric-like shifts of atoms V3, V4, V5, and V6 away from the 
centers of the surrounding oxygen octahedra. In contrast, the V5--V5 
dimerization and the strong ferroelectric-like shifts of atoms V1 and V2 
are present in both phases with only minor changes. Hence, even with 
respect to the details of the local octahedral distortions, the chain 
centers and ends behave similar to their parent structures $ {\rm VO_2} $ 
and $ {\rm V_2O_3} $. 

The LDA calculations were performed using the scalar-relativistic augmented 
spherical wave (ASW) method \cite{wkg} with a standard parametrization of 
the exchange-correlation potential \cite{vosko80}. Crystallographic data 
given by Canfield \cite{canfield90} were used. In order to 
account for the openness of the crystal structures, socalled empty spheres 
were included to model the correct shape of the crystal potential in large
voids. Optimal empty sphere positions and radii of all spheres were 
determined automatically \cite{vpop}. The basis set comprised V $ 4s $, 
$ 4p $, $ 3d $ and O $ 2s $, $ 2p $ as well as empty sphere states. 
Brillouin zone sampling was done using an increased number of $ {\bf k} $ 
points ranging from 108 to 2048 points within the irreducible wedge.

\begin{figure}
  \twoimages[width=65mm]{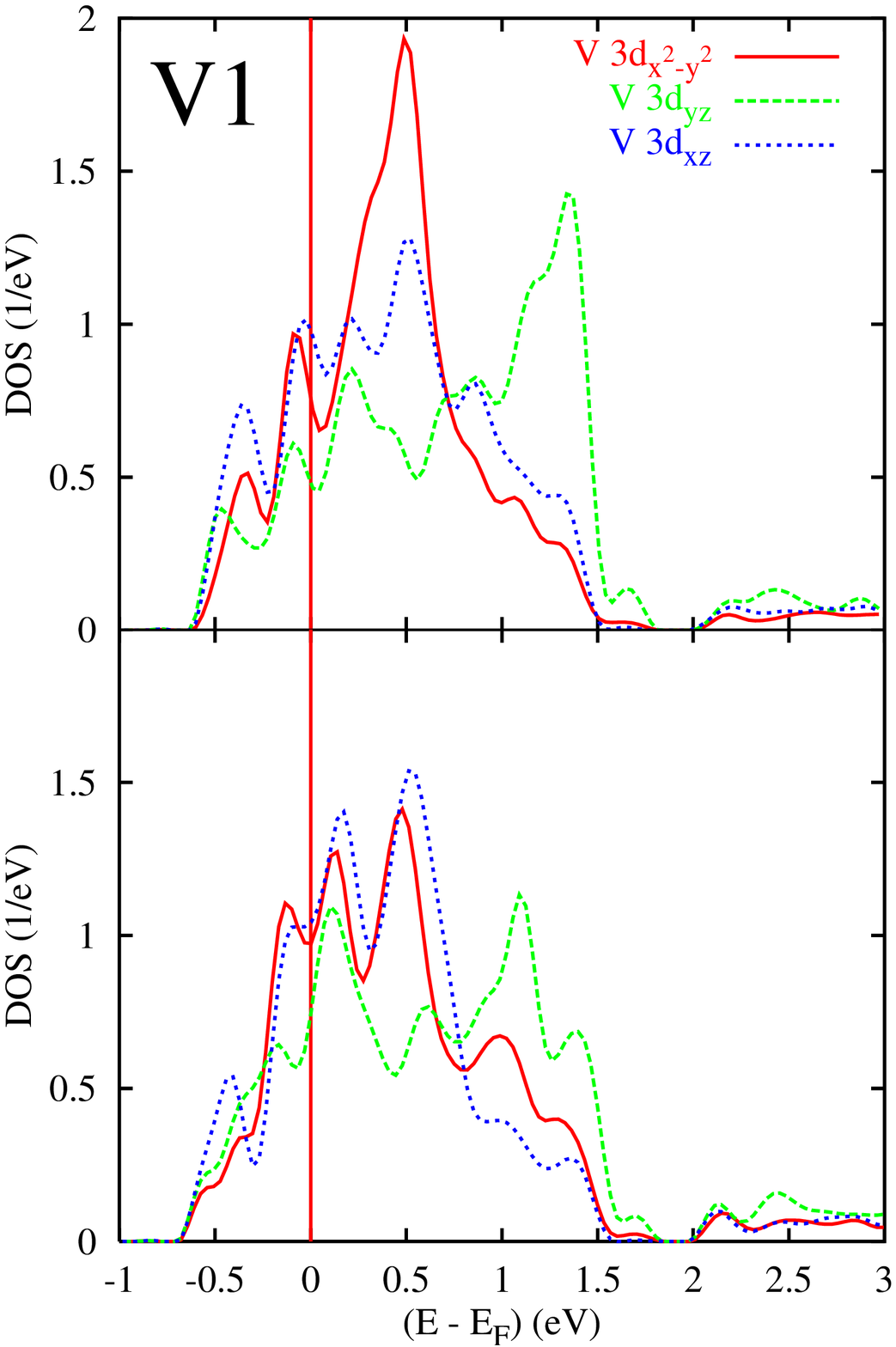}{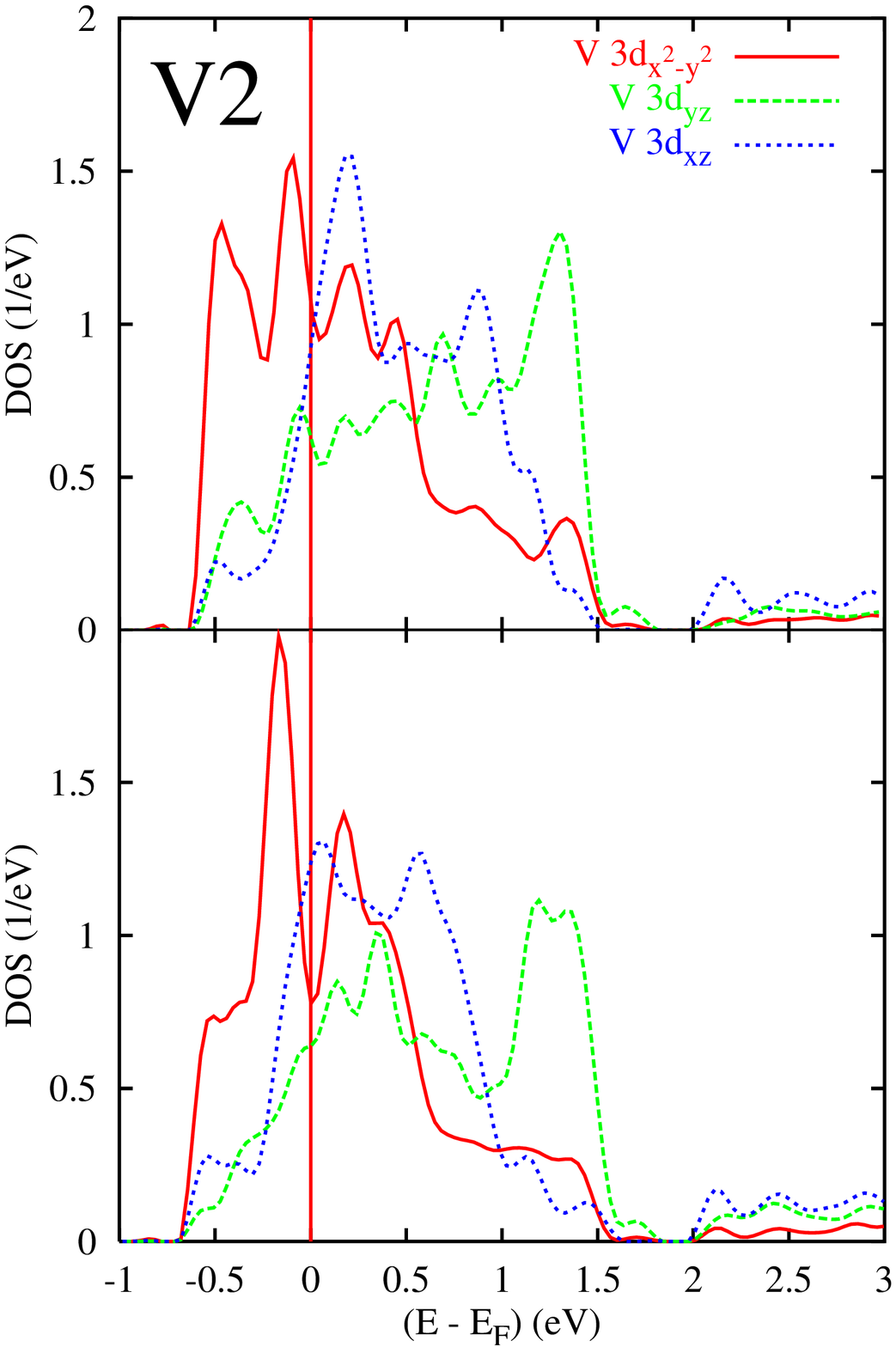}
  \caption{Partial V1 and V2 $ 3d $ DOS (per atom) of high- (top) and 
           low-temperature (bottom) $ {\rm V_6O_{11}} $.} 
  \label{fig2}
\end{figure}

\begin{figure}
  \twoimages[width=65mm]{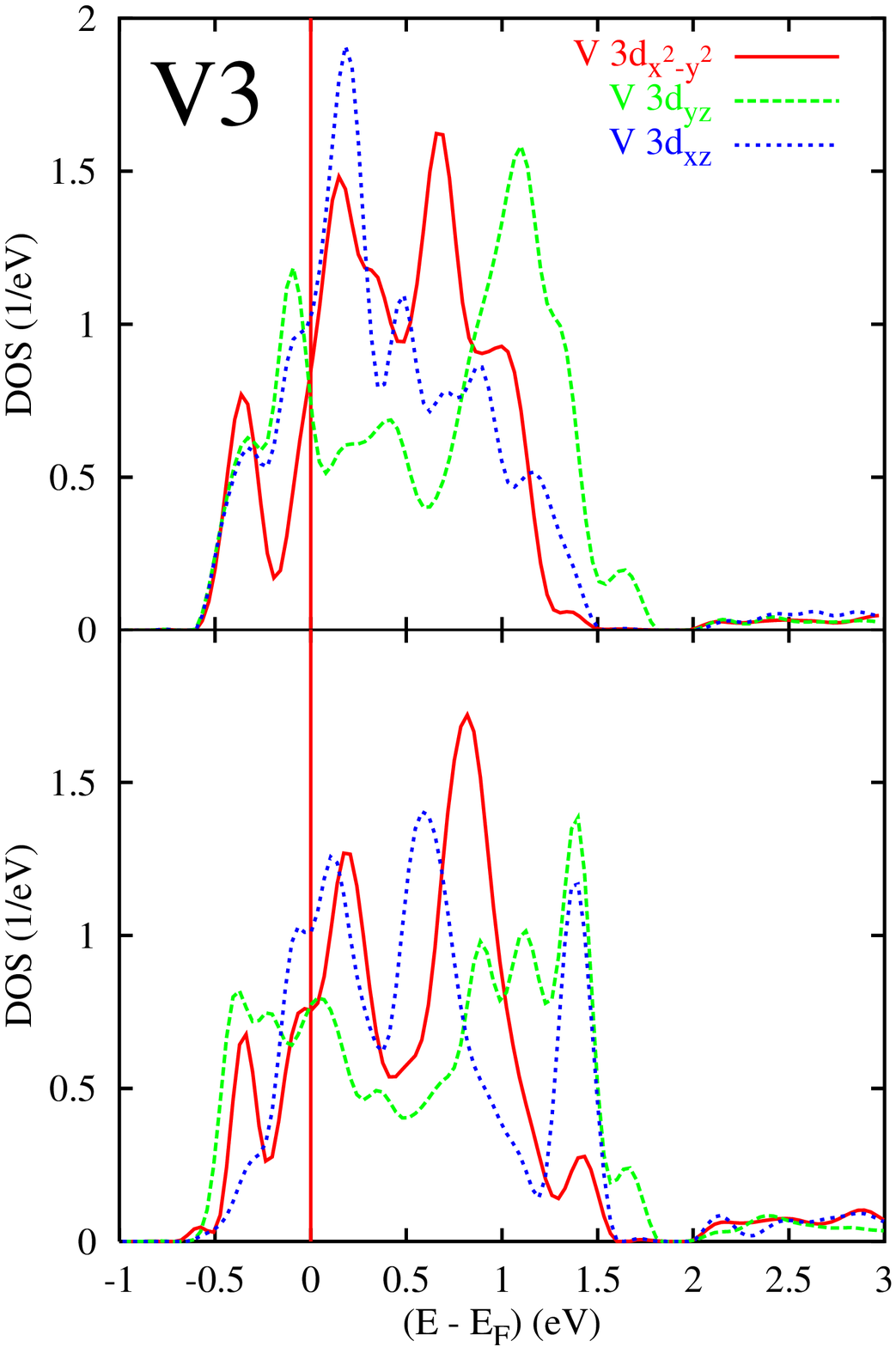}{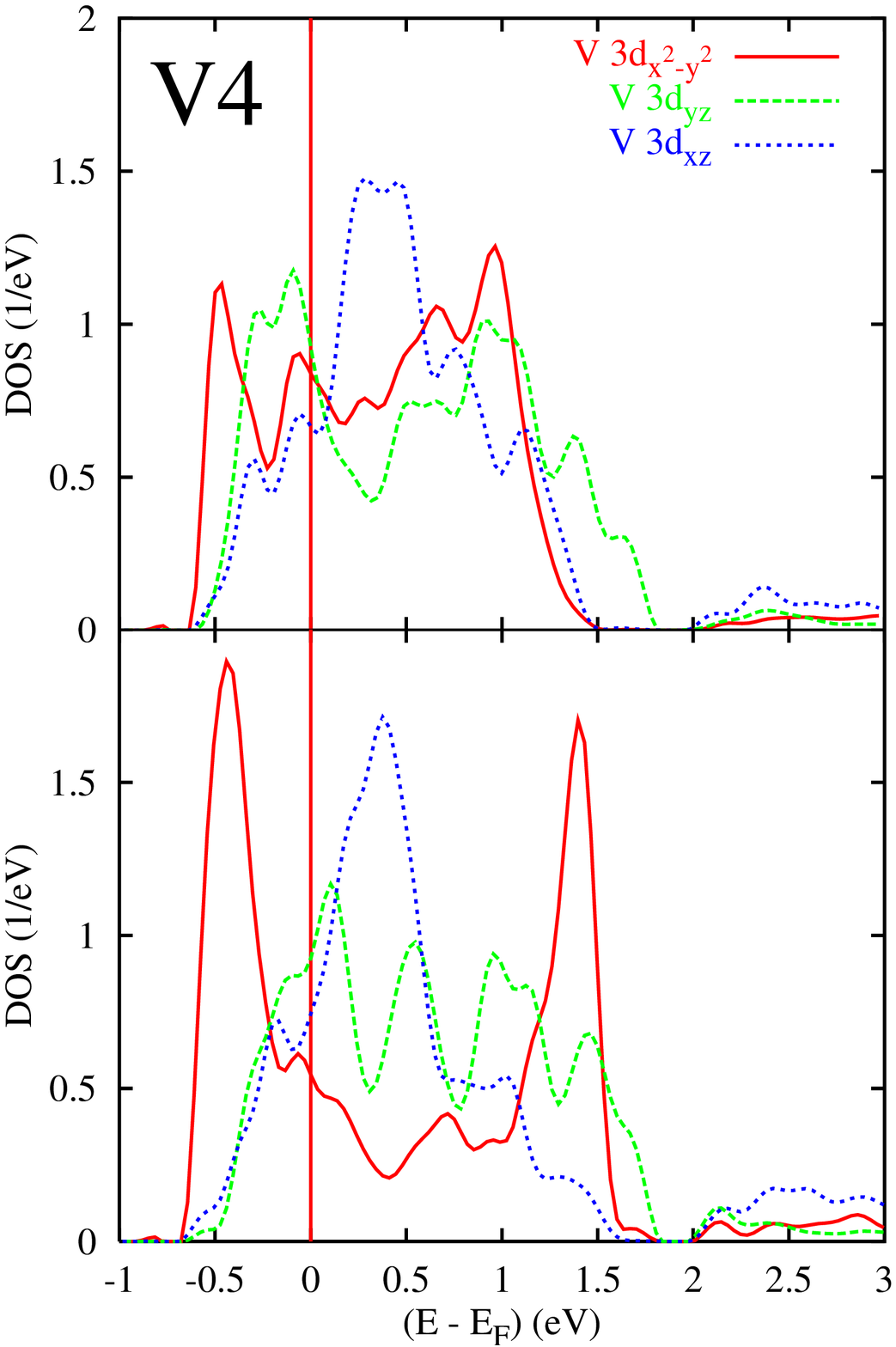}
  \caption{Partial V3 and V4 $ 3d $ DOS (per atom) of high- (top) and 
           low-temperature (bottom) $ {\rm V_6O_{11}} $.} 
  \label{fig3}
\end{figure}

Partial V $ 3d $ densities of states (DOS) resulting from calculations for 
the crystal structures of both phases are displayed in Figs.\ \ref{fig2} 
and \ref{fig3}. 
The gross features are very similar to those known from $ {\rm VO_2} $ 
and $ {\rm V_2O_3} $. Two groups of bands are identified, which result 
from crystal field splitting of the V $ 3d $ states due to the oxygen 
octahedra. Note that all orbitals are referred to the above local 
coordinate systems. While V $ 3d $ $ t_{2g} $ bands dominate in the 
energy region from $-0.7$ to 1.8\,eV, the $ e_g $ states (not included 
in Figs.\ \ref{fig2} and \ref{fig3}) are found mainly between 2.0 and 
4.8\,eV. Small contributions from the $ t_{2g} $ states in this energy 
region result from distortions of the octahedra. Not shown are O $ 2p $ 
states occupying the energy region from $-8.5$ to $-2.5$\,eV. 
Contributions of states in the energy range where the respective other 
orbital dominates are below 10\,\%, but still indicative of covalent 
bonding. The electronic states obtained for the low-temperature phase 
miss the insulating gap, what is due to the well known limitations of 
the LDA (and very similar to the situation reported for $ {\rm V_2O_3} $). 
However, these limitations are not important for the present work, as it 
aims primarily at the relation of the relevant orbitals to different 
local environments and their changes at the transition.

Going into more detail, we turn to the $ t_{2g} $ orbitals of those atoms which 
are involved in dioxide-like displacements, {\it i.e.}\ V4, V5, and V6. Their 
partial DOS are very similar among themselves and to those known from 
the rutile and monoclinic phase of $ {\rm VO_2} $ \cite{habil}. In 
particular, note the two peak structure of the high-temperature 
V4 $ d_{x^2-y^2} $ DOS, which is due to strong $ \sigma $-type metal-metal 
bonding parallel to the vanadium chains ($ c_{\rm prut} $). In the 
low-temperature structure, V4--V6 dimerization causes increased splitting 
of this DOS into bonding and antibonding branches. At the same time, both 
the $ d_{yz} $ and $ d_{xz} $ partial DOS experience an energetical upshift 
due to increased $ p $--$ d $ overlap arising from the ferroelectric 
displacement of the vanadium atom perpendicular to $ c_{\rm prut} $. As 
in $ {\rm VO_2} $ energetical separation between both types of bands is 
increased albeit not complete. The partial DOS of atom V5 display these 
features already in the metallic phase. In total, the partial DOS of the 
atoms V4, V5, and V6 can be fully understood in terms of the mechanisms 
known from $ {\rm VO_2} $. 

The situation is different for the atoms V1 and V2, which are in a 
sesquioxide-like environment. In particular, they are involved in 
metal-metal bonding via shared octahedral faces across the layers. 
Furthermore, these atoms do not experience any dimerization. As a 
consequence, 
except for small peaks and shoulders near $ -0.5 $ and between 1.0 to 
1.3\,eV, which are reminiscent of the respective peaks of the chain 
center atoms, the $ d_{x^2-y^2} $ DOS of atoms V1 and V2 consist of 
single rather broad peaks extending from $ -0.2 $ to 0.7\,eV. 
In both phases, subpeaks observed in this energy region do not have 
counterparts in the DOS of neighbouring atoms indicative of metal-metal 
bonding. Like the $ d_{xz} $ states of these atoms the $ d_{x^2-y^2} $ 
states may thus be regarded as rather localized. This is in contrast to 
the $ d_{yz} $ states, which show bonding-antibonding splitting due to 
the aforementioned overlap across the layers, {\it i.e.}\ parallel to 
$ a_{\rm prut} $. Since this direction is parallel to the hexagonal 
$ c_{\rm hex} $ axis of the corundum structure, the splitting of the 
$ d_{yz} $ states is equivalent to that of the $ a_{1g} $ states of 
$ {\rm V_2O_3} $. The partial DOS of atoms V1 and V2 thus can be interpreted 
in full accordance with the sesquioxide. 

Atom V3 is exceptional since it does not fit into the above schemes: 
Neither does this atom experience dimerization along the chains as in 
$ {\rm VO_2} $, nor does it have neighbouring vanadium atoms across the 
layers as in $ {\rm V_2O_3} $. Yet, the partial DOS are rather similar 
to those of atoms V1 and V2. The $ d_{x^2-y^2} $ and $ d_{xz} $ partial 
DOS form rather broad single peaks, the former still showing satellite 
peaks at $-0.4$ and 1.0/1.4\,eV reminiscent of the bonding and antibonding 
peaks of atom V5. In contrast, the $ d_{yz} $ partial DOS displays a 
pronounced double peak structure indicative of bonding-antibonding 
splitting. This puzzling situation is resolved by taking into account a 
different type of metal-metal bonding, namely that across octahedral faces 
along the sequences V1--V5--V3 and V2--V6--V4 parallel to $ b_{\rm prut} $. 
Hence, this $ \sigma $-type overlap and the resulting bonding-antibonding 
splitting of the $ d_{yz} $ states is present not only for atom V3, but for 
all atoms. It is also observed in $ {\rm VO_2} $ \cite{habil}. For the 
$ d_{yz} $ states of atoms V1 and V2 it adds to the above bonding parallel 
to $ a_{\rm prut} $; note that the respective antibonding peaks merge in 
the high-temperature phase. Although the coordination of atom V3 differs 
from that of all other atoms, its electronic properties are thus similar 
to that of the chain end atoms due to $ \sigma $-type metal-metal overlap 
via octahedral faces within the layers. Hence, metal-metal bonding across 
the layers is not the only source of the splitting of the $ d_{yz} $ states. 

In conclusion, while overlap of O $ 2p $ and V $ 3d $ states places the 
V $ 3d $ $ t_{2g} $ states near the Fermi energy, the detailed electronic 
properties of the latter and the MIT of $ {\rm V_6O_{11}} $ are strongly 
influenced by the {\em local} metal-metal coordination. For this reason, 
the partial DOS for each atom may be regarded as local quantities. Near 
the chain centers (atoms V4 to V6) dimerization and ferroelectric-like
displacements via strong electron-lattice interaction cause a splitting of
the $ d_{x^2-y^2} $ states and an energetical upshift of the $ d_{yz} $ and 
$ d_{xz} $ states. This is in complete analogy to the behaviour known 
from $ {\rm VO_2} $. 

In contrast, near the chain ends (atoms V1, V2, and V3) the $ d_{x^2-y^2} $ 
states experience less metal-metal overlap and, hence, become much more 
localized; their partial DOS resemble those of the $ d_{xz} $ states in the 
chain centers. As a consequence, electronic correlations might play a greater 
role. Being subject to metal-metal bonding only within the layers, the 
$ d_{yz} $ partial DOS of atom V3 displays bonding-antibonding splitting 
very similar to that of atoms V1 and V2. In addition, these atoms 
experience metal-metal overlap across the layers. We conclude that the 
V--V overlap within the layers is at least as important as the one
perpendicular. In summary, the MIT of $ {\rm V_6O_{11}} $ is interpreted as 
resulting from electron-lattice interaction and electronic correlations 
in the dioxide- and sesquioxide-like regions of the crystal, respectively. 

Our findings have important implications for the understanding of 
$ {\rm V_2O_3} $. According to the above results the electronic properties 
of this material are influenced by (i) the localization of the 
$ d_{x^2-y^2} $ states, and (ii) the splitting of the $ d_{yz} $ states due 
to metal-metal bonding within layers perpendicular to the hexagonal
$ c_{\rm hex} $ 
axis. Splitting of the $ a_{1g} $-like states due to overlap parallel to 
$ c_{\rm hex} $ might be less important than commonly assumed.  Both effects 
could not be analysed in previous studies of $ {\rm V_2O_3} $ due to 
the high symmetry of this material. Yet, they are accessible to our present 
analysis of the sesquioxide-like regions of $ {\rm V_6O_{11}} $. 
Nevertheless, the electronic structure as well as the  MIT of
$ {\rm V_6O_{11}} $ remain a crucial test case for all theories aiming at
a correct description of both $ {\rm VO_2} $ and $ {\rm V_2O_3} $.

\acknowledgments
Fruitful discussions with S.\ Horn and S.\ Klimm are gratefully 
acknowledged. This work was supported by the Deutsche 
Forschungsgemeinschaft through SFB 484.

\end{document}